\documentclass[twoside,twocolumn]{article}

\usepackage{blindtext} 

\usepackage[sc]{mathpazo} 
\usepackage[scale=0.93]{tgpagella} 
\usepackage[T1]{fontenc} 
\usepackage{microtype} 

\usepackage[english]{babel} 

\usepackage[hmarginratio=1:1,top=32mm,columnsep=20pt]{geometry} 
\usepackage[hang, small,labelfont=bf,up,textfont=it,up]{caption} 
\usepackage{float}
\usepackage{booktabs} 

\usepackage{lettrine} 

\usepackage{enumitem} 
\setlist[itemize]{noitemsep} 

\usepackage{abstract} 

\usepackage{titlesec} 
\renewcommand\thesection{\Roman{section}} 
\renewcommand\thesubsection{\roman{subsection}} 
\titleformat{\section}[block]{\large\scshape\centering}{\thesection.}{1em}{} 
\titleformat{\subsection}[block]{}{\thesubsection.}{1em}{} 

\usepackage{fancyhdr} 
\usepackage{titling} 

\usepackage[dvipsnames]{xcolor}
\usepackage{hyperref}

\colorlet{mylinkcolor}{Black}
\colorlet{mycitecolor}{Black}
\colorlet{myurlcolor}{Blue}
\usepackage{graphicx}
\usepackage{cite}

\hypersetup{
  linkcolor  = mylinkcolor,
  citecolor  = black,
  urlcolor   = myurlcolor,
  colorlinks = true,
}

\usepackage{graphicx}

\newcommand{\ORCIDDisplay}[1]{\href{http://orcid.org/\#1}{\includegraphics[scale=.10]{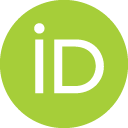}}}
\newcommand{\firstAuthor}{Gopal P. Sarma}
\newcommand{\firstAuthorEmail}{gopal.sarma@emory.edu}
\newcommand{\firstAuthorAffiliation}{School of Medicine, Emory University, Atlanta, GA USA}
\newcommand{\firstAuthorORCID}{0000-0002-9413-6202}

\pretitle{\begin{center}\huge\bfseries} 
\posttitle{\end{center}} 
\title{Doing Things Twice (Or Differently): Strategies to Identify Studies for Targeted Validation} 
\author{
\textsc{
\firstAuthor\ORCIDDisplay{\firstAuthorORCID}\textsuperscript{1}\thanks{Email: \firstAuthorEmail}\hspace{2pt} 
} \vspace{8pt} \\ 
\normalsize 1. \emph{\firstAuthorAffiliation}\\ 
}
\date{} 
\renewcommand{\maketitlehookd}{%
\begin{abstract}
\noindent The ``reproducibility crisis'' has been a highly visible source of scientific controversy and dispute.  Here, I propose and review several avenues for identifying and prioritizing research studies for the purpose of targeted validation.  Of the various proposals discussed, I identify scientific literature text mining as being a strategy that merits greater attention among those interested in reproducibility.  I argue that the tremendous potential of scientific literature text mining for uncovering high-value research studies is a significant and rarely discussed benefit of the transition to a fully open-access publishing model.  
\end{abstract}
}

\begin{document}

\maketitle


\section{Introduction}
In recent years, significant attention has been given to problems with reproducibility in many areas of science.  Some of these analyses have been theoretical in nature \cite{Ioannidis2005} while others have been focused efforts aimed at replicating large numbers of studies in a specific field \cite{Prinz2011, begley2012drug} . \\

The issue has become sufficiently high profile that it has been dubbed the ``reproducibility crisis,'' and is now a major topic of and debate in both the scientific \cite{gunn2014reproducibility, adam2002journals, check2005korean, Horton2015, Campbell2015a} and popular press \cite{economist_trouble, discover_neuroskeptic, nytimes, wired_crisis, wsj_reproducibility}.  \\

One thing is clear---we simply do not know what the ``reproducibility distribution'' looks like for the entirety of science.  Taking this position as a starting point, how then do we identify and prioritize published results to investigate in greater detail?  Should reproducibility initiatives be strictly local and originate from individual scientists themselves, or should there be more global, distributed efforts as well?  In this brief note, I examine this and other questions and propose several strategies for identifying key results to be the focus of validation efforts.  


\section{Uncovering ``Linchipin'' Results}
It goes without saying that all scientific results, even the true ones, are not created equally.  In order to use resources efficiently, we would ideally identify ``linchipin'' results, that is, those studies which would carry the highest impact if we had greater certainty in their outcome.  For instance, in cases where poorly conducted or fraudulent studies form the basis for guidelines or procedures in medicine, lengthy retractions can have significantly deleterious consequences for the public \cite{bouri2014meta}.  To use an ecological metaphor, these studies might be described as ``keystone species'' of the scientific ecosystem.  How can we uncover such results?  

\subsection{Reproduction Versus Validation}
Although the phrase ``reproducibility crisis'' has taken root in contemporary discussions, simply re-doing an experiment may not always be the most appropriate course of action.  For instance, there might be linchipin theoretical results which simply need greater scrutiny or investigation with alternative methods.  The same may be true of certain experimental results, where the highest value would be gained from re-thinking a given experimental design using alternative techniques. \\

Therefore, to broaden the scope of the discussion to include all scientific results, not just experimental ones, as well as approaches other than simply repeating the original study under question, I will use the phrase ``validation effort'' rather than ``reproducibility effort.''

\subsection{Have Individual Scientists Initiate Validation Efforts}
A purely ``local'' approach would be for validation efforts to be initiated by investigators themselves.  For instance, Schooler and several colleagues at UCLA arrived at an agreement whereby each researcher's lab would attempt to replicate the results of the others prior to publication \cite{schooler2014metascience}.  \\

However, not all scientists are in a position to create such arrangements.  Therefore, formal mechanisms for arranging validation efforts should be encouraged.  As an example, this is the principle behind ScienceExchange's Reproducibility Project\footnote{\url{http://validation.scienceexchange.com/}}, a marketplace for scientists to identify researchers from a network of laboratories to validate their research.  

\subsection{Polling Scientists or Crowdsourcing} 
This approach would be more globally oriented and could be initiated by funding agencies, individual laboratories, or by ``open science'' projects.  The strategy would be to distribute polls to scientists in different disciplines asking them what they believe to be high-value results.  Like any poll, a number of practical issues will arise in arriving at reliable data.  Questions will likely have to be written by scientists with sufficient experience in a given field to elicit reliable answers and to follow-up on ambiguities in responses.  The questions will have to be framed appropriately.  The administrators of such polls will have to be mindful of the fact that individuals could use a call for reproducibility as a political tactic for attacking a competitor's research. \\

Nonetheless, polling scientists is likely a very straightforward strategy to uncover high-value results.  The results of such polls could be used by funding agencies, independent foundations, or individual scientists themselves in deciding how to allocate resources for validation efforts.

\subsection{Let Larger Research Agendas be the Focal Points} 
This strategy would be to focus validation efforts around those results that form the foundation of major research agendas.  For example, while soliciting proposals for new programs, funding agencies could ask scientists to submit a bibliography containing results upon which their proposed research relies.  Or more directly, researchers could be asked to submit a separate document alongside their grant proposals suggesting experiments that merit additional investigation and which would advance their own research.  Scientists could also publish such documents independent of grant applications, perhaps along the lines of a review article.

\subsection{Scientific Literature Text Mining}
Scientific literature text mining refers to the use of data analytic techniques to treat the scientific corpus itself as a massive data set for analysis \cite{Sarma2017Scientific}.  Although the growth of data science has largely been driven by commercial applications in social media and business intelligence, we are now beginning to see the applications of data science to the scientific literature as well. \\

At the most basic level, article recommendation by Google Scholar or the many reference managers used by researchers is an example of data science applied to scientific papers.  Other examples of contemporary research in scientific literature text mining include fraud detection (applying natural language processing to uncover linguistic signatures of fraudulent research) \cite{markowitz2015linguistic}, characterizing the emergence of global scientific trends (using $n$-grams and patent citation networks to model the flow of ideas and technological development) \cite{solee2013evolutionary}, and resource allocation in the biomedical sciences (developing metrics which incorporate disease burden, research literature coverage, and clinical trial coverage to uncover underfunded areas of research) \cite{yao2015health}.  \\

One can easily imagine using techniques from machine learning and natural language processing to identify linchpin results, perhaps by examining citations networks, or using entity extraction to model the emergence of new terminology and concepts.  The process would not need to be fully automated.  We could employ a hybrid approach whereby data analytic techniques allow us to narrow down a corpus of tens of thousands of research papers to a few dozen or a hundred.  Subsequently, with the guidance of experts and manual curation, we would arrive at a list of candidate results or studies to be the focus of targeted validation efforts.

\section{Conclusion}
Modern science is witnessing many growing pains, one of which is an increasing concern about the quality of research, as measured by reproducibility, in a number of distinct areas of inquiry.  Although this concern has quickly grown to the point of being labeled a ``crisis,'' the reality is that we simply do not know what the distribution of reproducibility rates looks like for the entirety of science. \\

Nonetheless, this very uncertainty should be sufficient motivation to put into place procedures and incentives to increase the reliability of published results.  There are many approaches to take in addressing this problem, a few of which have been outlined above.  \\

Most of these ideas have been discussed or attempted in some form or another in recent years.  The proposal that has received the least attention in the context of the reproducibility crisis is scientific literature text mining.  One of the primary roadblocks to open-ended, exploratory data analysis with a large corpus of scientific papers is restrictions on their availability---in other words, closed access publishing models. Therefore, researchers should consider the applications of scientific literature text mining to uncovering linchpin results to be a key motivating factor in encouraging a complete transition to an open access publishing model.  

\section*{Acknowledgments}
I would like to thank Adam Safron, P. Ravi Sarma, and Daniel Weissman for insightful discussions and feedback on the manuscript.  A special thanks to A.S. for suggesting the phrase ``keystone species.''  

\section*{ORCID}
\makebox[2.5cm]{\firstAuthor} \raisebox{-.26\height}{\includegraphics[scale=.10]{orcid128}} \href{http://orcid.org/\firstAuthorORCID}{\firstAuthorORCID}\\

\bibliographystyle{ieeetr}
\bibliography{doing_things_twice}


\end{document}